\documentstyle[12pt,subeqn,epsf,fullpage]{article}
\def\be{\begin{equation}}
\def\ee{\end{equation}}
\def\bea{\begin{eqnarray}}
\def\eea{\end{eqnarray}}
\begin{document}
\begin{flushright}
TIFR/TH/99-40 
\end{flushright}
\bigskip\bigskip
\begin{center}
{\Large{\bf CONTEXTUAL DETERMINISTIC QUANTUM MECHANICS}}
\\[1cm]
{\large S.M. ROY} \\[1cm]
Tata Institute of Fundamental Research \\
Homi Bhabha Road, Mumbai 400 005, India \\[1cm]
E-mail: shasanka@theory.tifr.res.in
\end{center}
\bigskip\bigskip\bigskip

\noindent \underbar{Abstract.}  We present a simple proof of quantum
contextuality for a spinless particle with a one dimensional
configuration space.  We then discuss how the maximally realistic
deterministic quantum mechanics recently constructed by this author
and V. Singh can be applied to different contexts.

\vfill

\noindent PACS: 03.65

\newpage

\noindent 1. \underbar{Context Dependence of Quantum Probabilities.}
\medskip

A quantum state vector $|\psi\rangle$ specifies the probabilities
$|\langle \alpha|\psi\rangle|^2 =$ probability of observing the set of
eigenvalues $\{\alpha\} = \{\alpha_1,\alpha_2,\cdots,\alpha_n\}$ of a
complete commuting set of observables (CCS) $A =
\{A_1,A_2,\cdots,A_n\}$ in the experimental situation or `Context'
where $A$ is observed.  Equally, $|\psi\rangle$ specifies the
probabilities $|\langle \beta|\psi\rangle|^2$ for observing the
eigenvalues $\beta$ of a different CCS $B$ in the context where $B$ is
measured.  It is usually assumed that $A$ and $B$ cannot be measured
simultaneously if they contain any mutually noncommuting observables,
and hence that $|\langle \alpha|\psi\rangle|^2$ and $|\langle
\beta|\psi\rangle |^2$ refer to different contexts.

Thus, quantum probabilities are inherently context dependent.
Moreover, they cannot be embedded in context independent or
``classical'' probabilities.  This is the essence of Gleason's
theorem, Kochen-Specker theorem and Bell's theorem$^1$, which apply to
quantum systems with dimension of state vector space $\geq 3$.

An analogous question arises from a consideration of the Wigner
distribution function$^2$ $\rho(\vec x,\vec p)$.  It reproduces the
correct context dependent quantum probability densities for position
or momentum or for some components of position and commuting
components of momentum as different marginals of the same (context
independent) $\rho (\vec x,\vec p)$.  This is not a counter example to
the contextuality theorem just quoted since the Wigner function is not
positive definite and therefore cannot be given a probability
interpretation.  However, Cohen et al$^3$ have constructed positive
definite $\rho (\vec x,\vec p)$ whose marginals reproduce the quantum
position and momentum probability densities.  In a series of
investigations Roy and Singh$^4$ have gone much further; they
constructed `maximally realistic' causal quantum mechanics in which a
single positive definite $\rho(\vec x,\vec p)$ has marginals which
reproduce not just the quantum position and momentum probability
densities but also the quantum probability densities for $(n-1)$ other
complete commuting sets of observables, where $n$ is the dimension of
the configuration space.  Further, for $n \geq 2$, Martin and Roy$^5$
have proved a contextuality theorem: not all CCS can have their
quantum probability densitites reproduced as marginals of a single
positive definite phase space density.  It remains a conjecture$^4$
that $n+1$ is the maximum number of CCS whose probability densitites
can be reproduced as marginals of a single positive definite $\rho
(\vec x,\vec p)$.

We shall prove here that a contextuality theorem analogous to the
Martin-Roy theorem holds for $n=1$ too.  We shall also give a simple
proof of the Kochen-Specker theorem for a system of two spin half
particles which is equivalent to an earlier proof of Peres and
Mermin$^6$. 

We then discuss the Roy-Singh deterministic quantum mechanics for
$n=1$ allowing simultaneous reality of position and momentum
probability densities.  Equally well, any other pair of canonically
conjugate noncommuting variables can be simultaneously realized.
However, the contextuality manifests itself: different choices of the
simultaneously realised pair of dynamical variables lead to different
phase space densities.   The Roy-Singh deterministic quantum theory
which allows $(n+1)$ CCS to be simultaneously realized is much more
realistic than ordinary quantum mechanics or de Broglie-Bohm
deterministic quantum mechanics$^7$ (which allows only 1 CCS
viz. position to be realized).  It would be exciting to seek
experimental evidence of simultaneous realization of $(n+1)$ CCS in
one experimental setup.
\bigskip\bigskip\bigskip

\noindent 2. \underbar{A Simple Proof of Kochen-Specker theorem for a
System of Two Spin Half Particles.}
\medskip

Consider the possibility of a noncontextual microscopic theory
(`hidden variable theory'?) more detailed than ordinary quantum
mechanics in which a microstate assigns definite values for any set of
observables even if the set contains noncommuting observables of
quantum mechanics (which ordinarily refer to different contexts).  Can
such a theory be consistent with quantum predictions?

Quantum predictions imply that any commuting subset $A =
\{A_1,A_2,\cdots,A_n\}$ of operators can be simultaneously measured.
If the commuting operators obey the operator relation 
\be
f(A_1,A_2,\cdots,A_n) = 0
\label{one}
\ee
the assigned values $v(A_i)$ in the noncontextual theory must also
obey 
\be
f(v(A_1),v(A_2),\cdots,v(A_n)) = 0.
\label{two}
\ee
This is because the assigned set of values $\{v(A_i)\}$ must be an
allowed set of simultaneous eigenvalues of $\{A_1,\cdots,A_n\}$;
taking the expectation value of the operator equation (\ref{one}) in
the quantum state which has eigenvalues $v(A_i)$ for the operators
$A_i$ we obtain Eq. (\ref{two}).  For a system of two particles of
spin 1/2 we shall repeatedly use special cases of Eqns. (\ref{one})
and (\ref{two}), viz. if $[A,B] = 0$, then
\be
v(AB) = v(A) v(B), v(A+B) = v(A) + v(B).
\label{three}
\ee
Consider the operator relation involving the Pauli spin operators
$\sigma_{1x}$, $\sigma_{1y}$ for particle 1, and analogous operators
for particle 2,
\be
\left((\sigma_{1x} \sigma_{2y}) (\sigma_{1y} \sigma_{2x})\right) +
\left((\sigma_{1x} \sigma_{2x}) (\sigma_{1y} \sigma_{2y})\right) = 0
\label{four}
\ee
where each parenthesis encloses two commuting operators.  Hence,
applying (\ref{three}) repeatedly we obtain 
\be
2v(\sigma_{1x}) v(\sigma_{2y}) v(\sigma_{1y}) v(\sigma_{2x}) = 0.
\label{five}
\ee
But the left-hand side equals $\pm 2$ since the eigenvalue of a Pauli
matrix equals $\pm 1$.  The contradiction shows the impossibility of a
noncontextual extension of quantum mechanics, which is the
Kochen-Specker theorem.  This simple proof following immediately from
Eq. (\ref{four}) captures the essence of Mermin's earlier proof$^6$.

It is worth pointing out that the contradiction follows from two
hypotheses. 
\begin{enumerate}
\item[{(i)}] Simultaneous assignment of values to the operators
\be
\sigma_{1x} \sigma_{2x}, \sigma_{1x} \sigma_{2y}, \sigma_{1y}
\sigma_{2x}, \sigma_{1y} \sigma_{2y}, \sigma_{1z} \sigma_{2z},
\sigma_{1x}, \sigma_{1y}, \sigma_{2x}, \sigma_{2y}.
\label{six}
\ee
These assumptions are features also of Bell's theorem on violation of
Einstein locality.  It is not an additional assumption that the
products $(\sigma_{1x} \sigma_{2y}) (\sigma_{1y} \sigma_{2x}),
(\sigma_{1x} \sigma_{2x}) (\sigma_{1y} \sigma_{2y})$ (which
superficially involve factors \underbar{not} spacelike separated) have
values, since these products are equal to $\sigma_{1z} \sigma_{2z}$ and
$-\sigma_{1z} \sigma_{2z}$ respectively.
\item[{(ii)}] The values of any commuting subset of the operators
(\ref{six}) must be an allowed set of their simultaneous eigenvalues. 

In the cases of the subsets $\{\sigma_{1x} \sigma_{2y},\sigma_{1y}
\sigma_{2x}, \sigma_{1z} \sigma_{2z}\}$, $\{\sigma_{1x} \sigma_{2x},
\sigma_{1y} \sigma_{2y}, \sigma_{1z} \sigma_{2z}\}$ for which
commutation of the operators cannot be derived from a locality
postulate, this Kochen-Specker assumption is additional to Bell's
assumptions. 
\end{enumerate}
\bigskip\bigskip

\noindent 3. \underbar{Phase Space Contextuality Theorem For One
Dimensional Configuration Space.}
\medskip

Martin and Roy$^5$ have shown that in configuration space dimension
$\geq 2$, there cannot exist a positive definite phase space
distribution function whose marginals coincide with the quantum
position probability density, momentum probability density and joint
probability densities of commuting components of position and
momentum.  I shall prove here an analogous contextuality theorem for
configuration space dimension 1.

Let $X_{op} (\theta), P_{op} (\theta)$ be a continuum of
$\theta$-dependent operators defined by 
\be
\left(\matrix{X_{op} (\theta) \cr P_{op}
(\theta)/(m\omega)}\right) = \left(\matrix{\cos\theta & \sin\theta \cr
-\sin\theta & \cos\theta}\right) \left(\matrix{x_{op} \cr
p_{op}/(m\omega)}\right), 
\label{seven}
\ee
where $x_{op}$ and $p_{op}$ are the usual position and momentum
operators.  The constants $m,w$ have dimensions $[M]$, $[T^{-1}]$
respectively.  Then,
\be
[X_{op} (\theta),P_{op} (\theta)] = i\hbar
\label{eight}
\ee
We then prove the following contextuality theorem.
\bigskip

\noindent \underbar{Theorem.}  There exist state vectors
$|\psi\rangle$ for which it is impossible to find a positive definite
phase space density $\rho(x,p)$ with the property
\be
\int_{X(\theta)~{\rm fixed}} \rho(x,p) dP(\theta) = |\langle
X(\theta)|\psi\rangle|^2, ~{\rm for~all}~ \theta.
\label{nine}
\ee
Here $x,p,X(\theta),P(\theta)$ denote $c$-no. values belonging to the
spectrum of the operators $x_{op}$, $p_{op}$, $X_{op} (\theta)$,
$P_{op} (\theta)$, and the integral over $P(\theta)$ is done for a
fixed value of $X(\theta)$; i.e., we reexpress $\rho(x,p)$ in terms of
new variables $X(\theta), P(\theta)$,
\[
\rho(x,p) = \rho_\theta \left(X(\theta),P(\theta)\right)
\]
and then integrate over $P(\theta)$.
\bigskip

\noindent \underbar{Proof.} Consider the energy eigen function for the
first excited state of the oscillator,
\be
\langle x|\psi\rangle = {1 \over \sqrt{2}} \left({m\omega \over
\pi\hbar}\right)^{1/4} 2 \sqrt{m\omega \over \hbar} x \exp\left(-{1\over2}
{m\omega \over \hbar} x^2\right).
\label{ten}
\ee
The $X(\theta)$ representative of the same state vector $|\psi\rangle$
is given by a simple calculation
\[
\langle X(\theta)|\psi\rangle = \int dx \langle X(\theta)|x\rangle
\langle x|\psi\rangle,
\]
where $\langle X(\theta)|x\rangle$ is found from the differential eqn.
\bea
X(\theta) \langle x|X(\theta)\rangle &=& \langle x|\cos\theta x_{op} +
{\sin\theta \over m\omega} p_{op} |X(\theta)\rangle \nonumber \\[2mm]
&=& x \cos\theta \langle x|X(\theta)\rangle -i\hbar {\sin\theta \over
m\omega} {\partial \over \partial x} \langle x|X(\theta)\rangle. \nonumber
\eea
We thus find
\bea
\langle X(\theta)|\psi\rangle &=& \left({m\omega \over
\pi\hbar}\right)^{1/4} \sqrt{2m\omega \over \hbar} X(\theta)
\exp\Bigg[i \left({3\theta \over 2} - {\pi \over 4}\right) \nonumber
\\[2mm] && + {m\omega \over \hbar} \left({i\over2} \cot \theta
(X(\theta))^2 - {1\over2} (X(\theta))^2\right)\Bigg].
\label{eleven}
\eea
The most important point to notice is that
\be
\langle X(\theta)|\psi\rangle = 0 ~~{\rm at}~~ X(\theta) = 0
\label{twelve}
\ee
If $\exists$ positive definite $\rho(x,p)$ obeying (\ref{nine}), then
the equation 
\[
\int_{X(\theta)=0} dP(\theta) \rho(x,p) = 0
\]
implies that $\rho(x,p)$ must vanish identically on the line
\[
\cos\theta x + \sin\theta {p \over m\omega} = 0.
\]
For any given $x,p$ we can find $\theta$ which solves this equation.
Hence $\rho(x,p)$ must vanish identically for all $x,p$.  This is a
contradiction because Eq. (\ref{nine}) does not allow it.  Hence a
positive definite phase space density reproducing the quantum
marginals $|\langle X(\theta)|\psi\rangle|^2$ for all $\theta$ cannot
be found.

The method of proof is similar to that of Martin and Roy$^5$ except
for the fact that here I use canonical transformations which mix
position and momentum.
\bigskip

\noindent 4. \underbar{Contextual Deterministic Quantum Mechanics.} 
\medskip

De Broglie and Bohm$^7$ (dBB) constructed a deterministic quantum
mechanics in which only one CCS, namely position could have
(objective) reality independent of observations since only position
probability densities are correctly reproduced as a marginal of the
dBB phase space density,
\be
\rho_{dBB} = |\psi(\vec x,t)|^2 \delta(\vec p - \vec p_{dBB} (\vec
x,t)). 
\label{thirteen}
\ee
The individual trajectories given by a causal Hamiltonian are 
\be
\vec p_{dBB} = m {d\vec x \over dt} = \vec\nabla S(\vec x,t),
\label{fourteen}
\ee
where $\psi = R \exp(iS/\hbar)$, with $R$ and $S$ real; the ensemble
of trajectories ensure that at each time the quantum probability
density $|\psi(\vec x,t)|^2$ is reproduced.  It is easy to show
that$^8$
\[
\int \rho_{dBB} d\vec x \neq |\langle \vec p|\psi(t)\rangle|^2.
\]
In $n$ dimensional configuration space Roy and Singh$^4$ constructed a
new deterministic quantum mechanics in which the quantum probability
densities of $n+1$ CCS are simultaneously reproduced as marginals of
one positive definite phase space density.  The individual
trajectories are given by a causal Hamiltonian and hence the phase
space density is constant along the trajectory (Liouville property).
The new mechanics is conjectured to be maximally realistic in the
sense that it might be impossible to reproduce more than $n+1$ CCS as
marginals of a positive definite phase space density.  For $n=1$, two
simple phase space densities with this property given by Roy and Singh
are 
\be
\rho(x,p,t) = |\langle x|\psi\rangle|^2 |\langle p|\psi\rangle|^2
\delta\left(\int^p_{-\infty} dp'|\langle p'|\psi\rangle|^2 -
\int^{\epsilon x}_{-\infty} dx' |\langle \epsilon
x'|\psi\rangle|^2\right),
\label{fifteen}
\ee
where $\epsilon = \pm 1$, and we have written $|\psi\rangle =
|\psi(t)\rangle$ for brevity.  It is easy to check from this
expression which is symmetric between $x$ and $p$ that both $|\langle
x|\psi\rangle |^2$ and $|\langle p|\psi\rangle|^2$ are reproduced as
marginals. 

Nevertheless the mechanics is not context independent.  We already
know from the proof of the previous section that we cannot reproduce
$|\langle X(\theta)|\psi\rangle|^2$ for all $\theta$ as marginals.  In
spite of the fact that the operators $X_{op} (\theta)$ are in general
non-trivial linear combinations of the position and momentum variables
which may be somewhat unfamiliar, they are self-adjoint operators and
therefore observables.  For any given $\theta$, the Roy-Singh
procedure can be used to obtain a phase space density which reproduces
$|\langle X(\theta)|\psi\rangle|^2$ and $|\langle
P(\theta)|\psi\rangle|^2$ as marginals.  (Just replace $x,p$ by
$X(\theta), P(\theta)$ on the right-hand side of
Eq. (\ref{fifteen})).  The important point is that these phase space
densities are $\theta$-dependent and therefore must refer to different
contexts: each context refers to realizing $|\langle
X(\theta)|\psi\rangle|^2$ and $|\langle P(\theta)|\psi\rangle|^2$ for
a definite $\theta$.  Both the features of the Roy-Singh theory, (i)
realizing two probability densities instead of one in $dBB$ theory,
and (ii) the inevitable contextuality, are exhibited in the phase space
density.  We may mention that a $dBB$ like theory which reproduces
only one marginal, but with position replaced by a non-trivial linear
combination of position and momentum has been considered before in the
context of a quantum potential approach to a class of quantum
cosmological models$^9$.

For $n > 1$, it is even easier to exhibit the context dependence of
the Roy-Singh phase space densities.  E.g. for $n=2$, consider
\bea
\rho_\theta (\vec x,\vec p,t) &=& |\langle
x_1(\theta),x_2(\theta)|\psi\rangle|^2 |\langle
p_1(\theta),x_2(\theta)|\psi\rangle|^2 |\langle
p_1(\theta),p_2(\theta)|\psi\rangle|^2 \nonumber \\[2mm] 
&& \delta\left(\int^{p_1(\theta)}_{-\infty} |\langle
p'_1,x_2(\theta)|\psi\rangle|^2 dp'_1 - \int^{x_1(\theta)}_{-\infty}
|\langle x'_1,x_2(\theta)|\psi\rangle|^2 dx'_1\right) \nonumber
\\[2mm] && \delta\left(\int^{p_2(\theta)}_{-\infty} |\langle
p_1 (\theta), p_2'|\psi\rangle|^2 dp'_2 -
\int^{x_2(\theta)}_{-\infty} |\langle p_1(\theta),x'_2|\psi\rangle|^2
dx'_2\right), 
\label{sixteen}
\eea
where
\be
\left(\matrix{x_1(\theta) \cr x_2(\theta)}\right) =
\left(\matrix{\cos\theta & \sin\theta \cr -\sin\theta &
\cos\theta}\right) \left(\matrix{x_1 \cr x_2}\right),
\label{seventeen}
\ee
and a similar equation defines $p_1(\theta),p_2(\theta)$ in terms of
$p_1,p_2$.  The phase space densities depend on $\theta$; for each
$\theta$, the three indicated quantum probability densities $|\langle
x_1(\theta),x_2(\theta)|\psi\rangle|^2$, $|\langle
p_1(\theta),x_2(\theta)|\psi\rangle|^2$ and $|\langle
p_1(\theta),p_2(\theta)|\psi\rangle|^2$ are reproduced as marginals.
The causal Hamiltonians corresponding to Eqns. (\ref{fifteen}) and
(\ref{sixteen}) have also been calculated. 

The experimental realization of contexts in which some specific
$(n+1)$ CCS are simultaneously realized remains a challenge. 
\bigskip

\noindent \underbar{Acknowledgements.}  I am grateful to Deepak Kumar
and other organizers of the Conference on `Quantum Many Body Physics'
at the Jawaharlal Nehru University, New Delhi, March 5-7 for inviting
me to present these results.  It is a great pleasure to acknowledge
that the wide physics interests of R. Rajaraman (in whose honour this
meeting was held) has always been inspiring.  Many of the ideas
reported here have been developed in collaboration with Virendra
Singh. 
\bigskip

\noindent \underbar{References}
\medskip

\begin{enumerate}
\item[{1.}] A.M. Gleason, J. Math. \& Mech. \underbar{6}, 885 (1957);
S. Kochen and E.P. Specker, ibid \underbar{17}, 59 (1967); J.S. Bell,
Physics \underbar{1}, 195 (1964).
\item[{2.}] E. Wigner, Phys. Rev. \underbar{40}, 749 (1932).
\item[{3.}] L. Cohen and Y.I. Zaparovanny,
J. Math. Phys. \underbar{21}, 794 (1980); L. Cohen, ibid
\underbar{25}, 2402 (1984).
\item[{4.}] S.M. Roy and V. Singh, Mod. Phys. Lett. \underbar{A10},
709 (1995); S.M. Roy and V. Singh, Phys. Lett. \underbar{A255}, 201
(1999); S.M. Roy and V. Singh, in `Deterministic Quantum Mechanics in
One Dimension', p. 434, R. Cowsik (Ed.), Proceedings of International
Conference on Non-accelerator Particle Physics, 2-9 January, 1994,
Bangalore, World Scientific (1995); V. Singh, Pramana \underbar{49}, 5
(1997); S.M. Roy, Pramana \underbar{51}, 597 (1998).
\item[{5.}] A. Martin and S.M. Roy, Phys. Lett. \underbar{B350}, 66
(1995). 
\item[{6.}] A. Peres, Phys. Lett. \underbar{A151}, 107 (1990);
N.D. Mermin, Phys. Rev. Lett. \underbar{65}, 3373 (1990).
\item[{7.}] L. de Broglie, Non-linear Wave Mechanics, A Causal
Interpretation, Elsevier, 1960; D. Bohm, Phys. Rev. \underbar{85},
166, 180 (1952); P.R. Holland, The Quantum Theory of Motion, Cambridge
University Press 1993 and Foundations of Physics \underbar{28}, 881
(1998); D. Bohm and B.J. Hiley, The Undivided Universe, Routledge,
London, 1993; E. Deotto and G.C. Ghirardi, Foundations of Physics
\underbar{28}, 1 (1998). 
\item[{8.}] T. Takabayasi, Prog. Theor. Phys. \underbar{8}, 143
(1952). 
\item[{9.}] A. Blaut and J. Kowalski-Glikman,
Class. Quant. Grav. \underbar{13}, 39 (1996).
\end{enumerate}

\end{document}